%% file: main.tex
\documentclass[
preprint,
showpacs,preprintnumbers,
bibnotes,
 amsmath,amssymb,
 aps,
 pra,
superscriptaddress,
longbibliography,
]{revtex4-2}

\usepackage{graphicx}
\usepackage[english]{babel}
\usepackage[utf8]{inputenc}
\usepackage{geometry}
\usepackage{amsmath,amsthm,amsfonts,amssymb,amscd,mathtools}
\usepackage{hyperref}
\usepackage{enumerate}
\usepackage{textcomp}
\usepackage{fixmath}
\usepackage{xcolor}

\usepackage{bibunits}

\usepackage{setspace}

\begin{document}

\title{\textbf{Asymmetric high-harmonic generation from subwavelength bianisotropic resonators}}

\author{Albert Mathew}
\thanks{These authors contributed equally to this work.}
\affiliation{Nonlinear Physics Centre, Research School of Physics, Australian National University, Canberra ACT 2601 Australia}

\author{Piyush Jangid}
\thanks{These authors contributed equally to this work.}
\affiliation{Nonlinear Physics Centre, Research School of Physics, Australian National University, Canberra ACT 2601 Australia}

\author{Rebecca Aschwanden}
\thanks{These authors contributed equally to this work.}
\affiliation{Institute for Photonic Quantum Systems (PhoQS) and Department of Physics,
Paderborn University, Warburger Straße 100, 33098, Paderborn, Germany}

\author{Yves Köppeler}
\affiliation{Institute for Photonic Quantum Systems (PhoQS) and Department of Physics,
Paderborn University, Warburger Straße 100, 33098, Paderborn, Germany}

\author{Thomas Zentgraf}
\email{thomas.zentgraf@uni-paderborn.de}
\affiliation{Institute for Photonic Quantum Systems (PhoQS) and Department of Physics,
Paderborn University, Warburger Straße 100, 33098, Paderborn, Germany}

\author{Sergey Kruk}
\email{sergey.kruk@tuni.fi}
\affiliation{Photonics Laboratory, Physics Unit, Tampere University, Tampere, Finland}

\begin{abstract}

High-harmonic generation (HHG) enables attosecond light pulses and table-top sources of coherent extreme-ultraviolet and soft X-ray radiation. Although HHG has long been associated with gases and plasma, nanostructured solids are emerging as new alternative sources enabling both the enhancement and control of HHG. 
Here, we experimentally demonstrate and theoretically describe that a single dielectric subwavelength resonator can act as a direction-selective high-harmonic source, enabling control over multiple harmonic orders through the excitation and hybridization of Mie modes.
The resonator's geometrical volume is $0.12 \lambda^3$, and its optical mode volume is $0.03 \lambda^3$ at its pump wavelength.
Structural asymmetry of the resonator along the propagation direction translates into different mode coupling under opposite illumination directions, resulting in pronounced forward-backward asymmetry in the generation of the third, fifth, and seventh harmonics.  
These results establish bianisotropic subwavelength resonators as a platform for flexible asymmetric generation of high harmonics, expanding the toolbox for controlling strong-field light-matter interactions with Mie-resonant nanophotonics.

\end{abstract}

\maketitle

\section{Introduction}

Observations of high-harmonic generation (HHG) from gases \cite{Ferray1988} and plasmas \cite{Burnett1977} were made in the decades following the invention of the laser. HHG from solids was observed more recently \cite{Ghimire2011,Goulielmakis2022,Ghimire2018}. All-solid-state HHG attracted attention by bringing new physics to strong-field light-matter interactions, such as non-centrosymmetric materials associated with the generation of even-order harmonics \cite{Schubert2014,NatComm2023_HHG,Shcherbakov2021}, at the same time promising smaller and simpler HHG systems.  However, bulk solids pose challenges for some of the key applications of HHG, including attosecond pulses and ultrashort-wavelength radiation due to strong dispersion and short-wavelength material absorption of solids.

The limitations of bulk solids open a unique opportunity for HHG in ultra-thin, nanostructured solids: nanoresonators and metasurfaces \cite{Zalogina2023,Vampa2017,Zograf2022,Shcherbakov2021,An2021,Abbing2022,Jangid2024}. The ultra-thin form factor mitigates the disadvantages of bulk solids, while careful engineering of subwavelength resonant elements enables both the enhancement and control of HHG \cite{Zubyuk2021} often via excitation of localized Mie modes or bound states in the continuum.

Enhancement of HHG has been investigated across various resonant nanophotonic platforms, beginning with plasmonic nanostructures
\cite{Vampa2017,Kim2008,Sivis2013,Sivis2017,Park2011,Huttner2021,You2023}. While plasmonic structures offered strong field enhancements combined with tight field confinements, they suffered from intrinsic Ohmic losses, resulting in relatively low laser damage thresholds. All-dielectric metasurfaces and individual dielectric subwavelength resonators came as an appealing alternative, providing optical responses with sharp resonances, small mode volumes, while at the same time withstanding higher laser powers \cite{Shcherbakov2021,Abbing2022,Zograf2022,Zalogina2023,Zograf2022,Zalogina2023, tonkaev2024even}. Capabilities of nanostructured solids in controlling the emission of HHG directly at the generation stage were demonstrated for wavefront control \cite{Abbing2022} and chirality \cite{Jangid2024}.

In this work, we demonstrate asymmetric control over the excitation and emission of high harmonics. In the recent past, arrays of nonlinear nanoresonators -- metasurfaces -- with engineered geometrical asymmetries achieved strong nonlinear nonreciprocal self-action \cite{Mathew2025_ENZ,Tripathi2024_VO2,King2024,Cotrufo2024}
as well as asymmetric generation of low-order 3$^{rd}$ \cite{Kruk2022} and 2$^{nd}$ \cite{Boroviks2023} harmonics.
Here, by leveraging the interplay between nonlinear light-matter interactions and bianisotropic response in asymmetric nanoresonators, we experimentally demonstrate direction-dependent generation of multiple harmonics. Our platform consists of individual subwavelength resonators composed of two layers: amorphous silicon (Si) and silicon nitride (Si$_3$N$_4$). At the excitation wavelengths, the resonant response is dominated by the dipolar and quadrupolar Mie modes of both the electric and the magnetic types. We show that different directions of excitation of such resonators lead to different multipolar compositions of their responses, which translates into different efficiencies in the generation of optical harmonics. Our demonstrations add engineered anisotropy of nanostructured solids as a new tool for controlling HHG with nanotechnology.

\begin{figure*}[htb]
    \centering
    \includegraphics[width=1\textwidth]{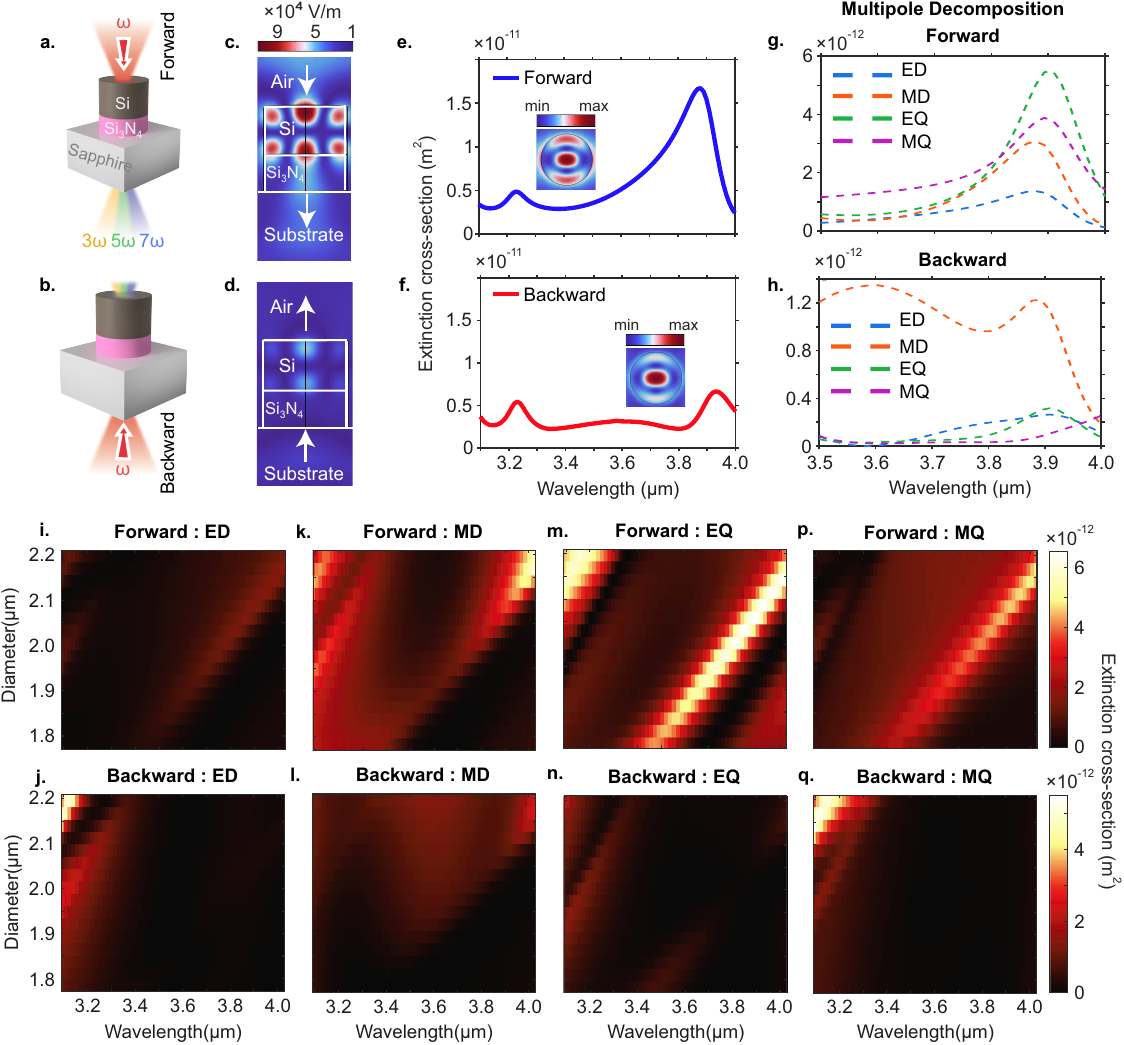}
    \caption{\textbf{Calculations of linear optical response of a Si-Si$_3$N$_4$ resonator.} \textbf{(a,b)} Concept of asymmetric generation of odd-harmonics, including the third ($3\omega$), fifth ($5\omega$), and seventh ($7\omega$) harmonics, from the subwavelength resonator under the ``forward'' and ``backward'' excitations. 
    \textbf{(c,d)} COMSOL calculations of near-field distribution within the resonator for the ``forward'' and ``backward'' excitations. Resonator diameter: 2.1 $\mu$m, Si layer height 1.21 $\mu$m and Si$_3$N$_4$ layer height 0.9 $\mu$m, excitation wavelength: 3875 nm. 
    \textbf{(e,f)} Total extinction cross-sections for the ``forward'' and ``backward'' excitations. Inset shows the horizontal cross section of near field distribution of electric field
    \textbf{(g,h)} Multipolar decompositions of scattering for the ``forward'' and ``backward'' excitations. ED, MD, EQ and MQ: electric dipole, magnetic dipole, electric quadrupole and magnetic quadrupole respectively.
    \textbf{(i,j,k,l,m,n,p,q)} Multipolar contribution of the effective scattering cross section as a function of the wavelength and the radius of the subwavelength resonator.
    }
    \label{fig:concept}
\end{figure*}


\section{Theory}

Subwavelength resonators used here are depicted in Figs.~\ref{fig:concept}a,b. The resonators are cylindrical in shape and consist of two layers -- amorphous silicon and silicon nitride -- deposited on a sapphire substrate. 

This bilayer design introduces structural asymmetry along the propagation direction, leading to bianisotropy \cite{Asadchy2018Bianisotropic,alaee2015all}. We note that here the bianisotropy arises from different refractive indices of Si and Si$_3$N$_4$ disks, otherwise identical in diameter. The bianisotropy is thus introduced via asymmetry of the refractive index distribution rather than geometrical asymmetry\cite{alaee2015all}.  We first calculate the linear response of a single, stand-alone subwavelength resonator in the mid-IR spectral range using COMSOL Multiphysics. We consider a cylindrical resonator of diameter 2.1 $\mu$m with the Si layer of height 1.21 $\mu$m and Si$_3$N$_4$ layer of height 0.9 $\mu$m. The computational domain is enclosed by perfectly matched layers (PMLs) that surround both the resonator and the substrate.  

Figs.~\ref{fig:concept}c,d show the corresponding near-field distributions at 3875~nm for the ``forward'' and ``backward'' directions. Here we define "forward" as illumination from the air side and "backward" as illumination from the substrate side.  Under forward illumination in the provided example, the field is markedly stronger and more tightly confined within the Si disk compared to backward illumination, with a maximum forward-to-backward field density ratio of up to 6.75. 

We determined the mode volume using finite element simulations performed in COMSOL Multiphysics. The effective mode volume was calculated using

\begin{equation}
V_{\mathrm{mode}} = 
\frac{\int \varepsilon(\mathbf{r}) |\mathbf{E}(\mathbf{r})|^2 \, dV}
{\max\left[\varepsilon(\mathbf{r}) |\mathbf{E}(\mathbf{r})|^2\right]},
\end{equation}

where $\varepsilon(\mathbf{r})$ is the dielectric permittivity of the material and $\mathbf{E}(\mathbf{r})$ represents the electric field distribution of the resonant mode across the resonator \cite{alaeian2020cavity,srinivasan2006cavity}. The quantity $\varepsilon(\mathbf{r}) |\mathbf{E}(\mathbf{r})|^2$ accounts for the time averaged electromagnetic energy density. In the equation, the numerator represents the total stored electromagnetic energy obtained by integrating the energy density over the resonator and the surrounding near-field region where the electric field is localized.

The calculated mode volumes are reported in dimensionless units normalized to $\lambda^3$ where $\lambda$ -- excitation wavelength in air. For a scenario in Fig.~\ref{fig:concept}c (resonator diameter of 2.1~$\mu$m excited at a pump wavelength of $\lambda=$3875~nm in forward direction), we obtained the mode volume as $0.03\lambda^3$, indicating strong subwavelength confinement of the electromagnetic field.

We proceed with calculations of the far-field response of the resonator. The far-field is obtained by first computing the background field without the resonator, and then using this field as the incident field for the scattering simulation with the resonator present. We calculate scattering, absorption and total extinction cross-sections of the resonator on the substrate for the two opposite excitations. 
The scattering cross-section, $\sigma_{\mathrm{sca}}$, is calculated from the closed-surface flux of the scattered Poynting vector as
\begin{equation}
\sigma_{\mathrm{sca}} = \frac{1}{I_0} \int_{A} \left( \mathbf{S}_{\mathrm{sca}} \cdot \mathbf{n} \right)\, dA
\end{equation}
where $\mathbf{n}$ is the outward normal vector, $\mathbf{S}_{\mathrm{sca}}$ is the scattered Poynting vector, and $I_0$ is the incident pump intensity taken as 100 W/cm$^2$. Next, the absorption cross-section, $\sigma_{\mathrm{abs}}$, is obtained by integrating the power-loss density $\Gamma$ over the resonator volume $V$ as
\begin{equation}
\sigma_{\mathrm{abs}} = \frac{1}{I_0} \int_{V} \Gamma \, dV
\end{equation}
The total electromagnetic power dissipation density is expressed as $\Gamma = \Gamma_{rh} + \Gamma_{ml}$, where the dielectric losses are given by
\begin{equation}
\Gamma_{rh} = \frac{1}{2}\,\mathrm{Re}\!\left(\mathbf{J}\cdot\mathbf{E}^{*}\right),
\end{equation}
and the magnetic losses are given by
\begin{equation}
\Gamma_{ml} = \frac{1}{2}\,\mathrm{Re}\!\left(i\omega\,\mathbf{B}\cdot\mathbf{H}^{*}\right),
\end{equation}
with $\mathbf{J} = \bar{\bar{\sigma}}\cdot\mathbf{E} + i\omega\mathbf{D}$ the total induced current density. These quantities are computed and integrated over the resonator volume $V$ to obtain $\sigma_{\text{abs}}$.

This yields the extinction cross-section, $\sigma_{\mathrm{ext}}$, given by
\begin{equation}
\sigma_{\mathrm{ext}} = \sigma_{\mathrm{sca}}+\sigma_{\mathrm{abs}}
\end{equation}
%
%

To evaluate the multipolar contributions to the scattering response, we employed exact multipole formulation as in~\cite{OptCommMultipole2018}. In this formalism, the effective total scattering cross section is decomposed into electric and magnetic dipole and quadrupole contributions derived from the polarization current density. Although the exact multipole decomposition is formally derived for a scatterer in a homogeneous background, our structure resides on a Sapphire substrate. In this setting, the reflectance at the resonator-substrate interface is small. Therefore, we consider the multipole decomposition as a qualitative diagnostic of the dominant current distribution. The presence of substrate means that the decomposition should be interpreted as approximate.


Our resonator lacks spatial inversion symmetry along its propagation axis, placing it strictly within the \(C_{\infty v}\) point symmetry group as per e.g. ref \cite{gladyshev2020symmetry}. From the near field distribution of the horizontal cross section of the nanoresonator (Fig.\ref{fig:concept}e,f inset)  we can infer that the field profile exhibits a distinct two-lobed structure split by a straight nodal line, which serves as the visual fingerprint of an \(m=1\) azimuthal configuration and thus the system matches the symmetric properties of $E_1$ subspace ($m=1$) \cite{pascale2019full}. Under the rules of the $C_{\infty v} $ group, the absence of inversion symmetry relaxes the parity based separation of multipolar channels observed in centrosymmetric systems. Since the system follows the symmetric properties of $E_1$ irreducible representation, its electric field directly couples to and cross-excites higher order multipoles of both electric and magnetic type. This transformation acts as a fundamental driving mechanism for strong multipolar bianisotropy \cite{bernal2014underpinning,mun2019importance,proust2016optical}.

The absence of inversion symmetry makes the two incidence directions physically distinct and each direction excites a different set of multipoles\cite{poleva2023multipolar}. Within the T-matrix formalism\cite{cannata2007scattering}, in centrosymmetric system, the parity operator commutes with the T matrix in the basis of vector spherical harmonics. But when the inversion symmetry is lifted in the $C_{\infty v}$ resonator, this constraint is lifted and the T-matrix acquires symmetry-allowed off-diagonal couplings 
and results in different multipolar scattering amplitudes for the opposite directions of excitation (Figs.~\ref{fig:concept}g,h). In particular, the ``forward'' excitation leads to efficient coupling to the quadrupole Mie modes -- both magnetic and electric (Fig.~\ref{fig:concept}g). The ``backward'' excitation couples most efficiently to the magnetic dipole mode (Fig.~\ref{fig:concept}h). The quadrupolar modes, in comparison to dipole modes, are associated with a higher Q-factor and tighter near-field confinements. Thus, when driven by an intense laser field, the nonlinear response in this case for the ``forward'' direction is expected to be stronger than the ``backward'' excitation scenario. We find this mechanism similar and complementary to the metasurfaces demonstrated previously in Ref.~\cite{Kruk2022}, in which a dipolar bianisotropic response led to an exchange of energy between the magnetic and electric Mie dipoles, whereas our system scales this phenomenon into the quadrupolar regime and extends its applications to high harmonic generation.

Fig.\ref{fig:concept}i-q shows the diameter-dependent multipole contributions of the nanoresonator. In forward direction we can see that there is sharp resonance contribution from the multipoles red-shifting as we increase the diameter of the resonator. Here quadrupole contribution dominates over the other multipoles. This coexistence of multipoles shows that they are not independent and they are tied to the same mathematical subspace -- the \(E_{1}\) irreducible representation. In backward direction the MD dominates over the other multipoles. Other contributions are significantly low compared to the forward direction. The dominant MD in backward direction unlike the dominant nature of EQ moment in forward direction itself is the evidence of multipolar bianisotropy in the nanoresonator.




We further examine the effect of the resonator diameter versus excitation wavelengths. Fig.~\ref{fig:hhg_maps}a shows the calculated extinction cross-section as a function of pump wavelength and resonator diameter (we keep the thickness of Si and Si$_3$N$_4$ layers fixed) for both illumination directions. The associated contrast map in the lower panel reveals a pronounced asymmetry between forward and backward responses. 


\section{Experiment}

\begin{figure*}[htb]
    \centering
    \includegraphics[width=1\textwidth]{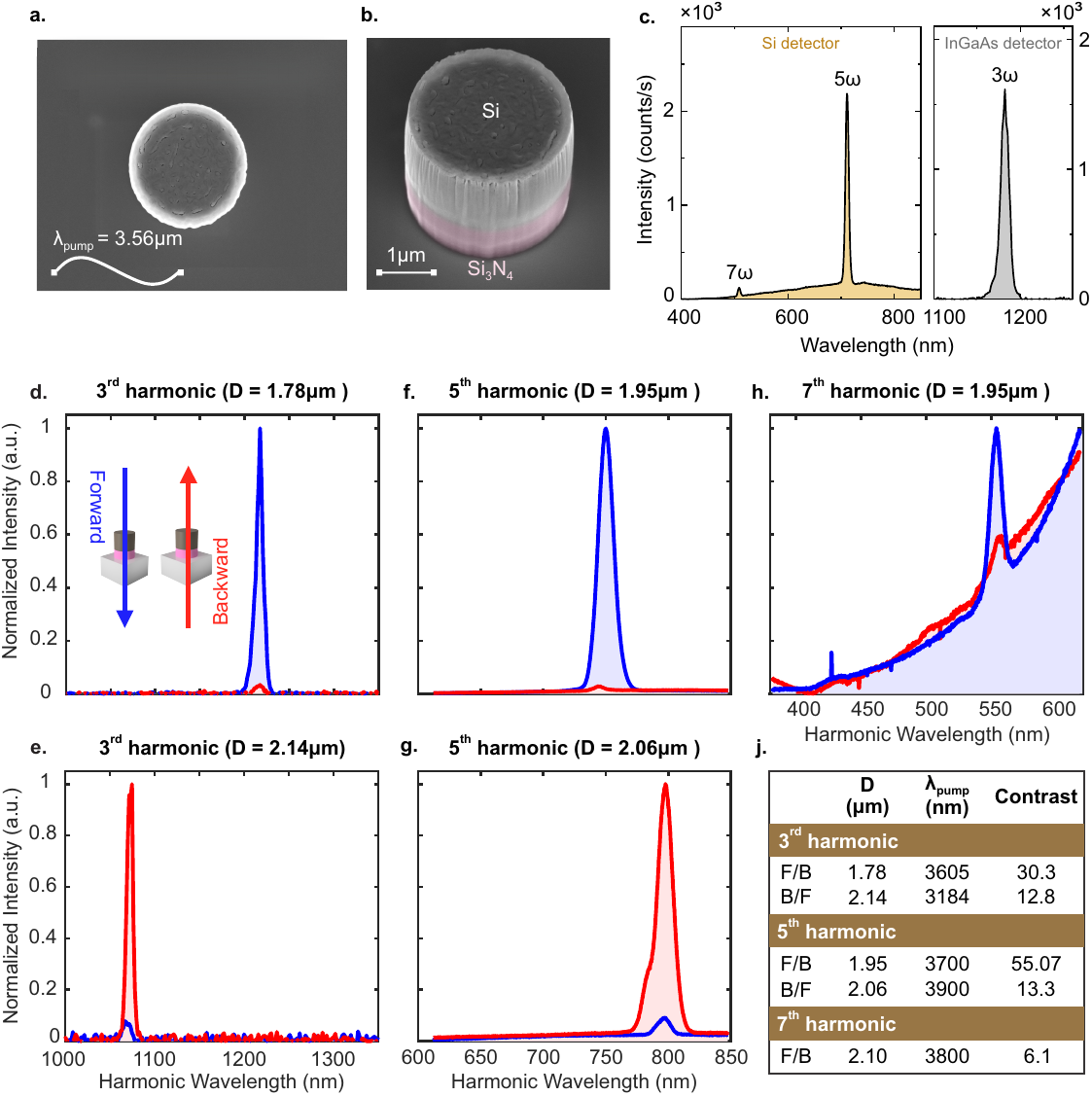}
    \caption{\textbf{Experimental results on asymmetric generation of multiple harmonics.} 
    \textbf{(a,b)} Scanning electron micrograph of one of the fabricated resonators of diameter 2.1$\mu$m: top view and side view. \textbf{(c)} The third (3$\omega$), fifth ($5\omega$) and seventh ($7\omega$) harmonics generation observed from the subwavelength resonator with diameter of 2.1 $\mu$m shown in \textbf{(a,b)} at an excitation wavelength of 3560~nm. The third harmonic data were collected using the spectrometer with an InGaAs-based detector, while the fifth and seventh harmonic data were collected using the Si-based spectrometer.
    \textbf{(d-h)} Examples of high forward-to-backward contrast (\textbf{d,f,h}) and high backward-to-forward contrast (\textbf{e,g}) for the third harmonic (\textbf{d,e}), the fifth harmonic (\textbf{f,g}) and the seventh harmonic (\textbf{h}). Corresponding resonator diameters and excitation wavelengths are indicated for each case. 
    \textbf{(j)} Table summary of experimentally measured forward-backward contrasts for the third, fifth and seventh harmonics.
    }
    \label{fig:SEM}
\end{figure*}

We fabricate subwavelength resonators consisting of an amorphous silicon and silicon nitride bilayer on a sapphire substrate. The Si$_3$N$_4$ and Si films are deposited using plasma-enhanced chemical vapour deposition. Silicon is deposited using only a 2\% silane-98\% argon gas mixture, while for the deposition of Si$_3$N$_4$ additionally NH$_3$ is added. Subsequently, a layer of polymethyl methacrylate (PMMA) electron-beam resist is spin-coated onto the sample, and the cylindrical patterns of subwavelength resonators of varying diameters are defined with electron-beam lithography. We keep large spacings between the individual resonators on the order of 28 $\mu$m, which is sufficient for each resonator to be excited individually in subsequent optical experiments with a focused laser beam.
After development in methylisobutylketone (MIBK), a 45 nm thick chromium (Cr) hard mask is deposited using electron-beam evaporation. The lift-off process in hot acetone followed by dimethyl sulfide (DMS) removes the resist, leaving the patterned Cr mask on the bilayer film. The pattern is then transferred into the underlying materials by sequential reactive-ion etching of the a-Si layer (with SF$_6$ and C$_4$F$_8$) followed by the Si$_3$N$_4$ layer (with CHF$_3$ and O$_2$). Finally, the residual Cr mask is removed using a wet etchant based on ceric ammonium nitrate and perchloric acid. Figs.~\ref{fig:SEM}a,b show the scanning electron micrograph of a fabricated bilayer cylindrical resonator.
 
Next, we perform nonlinear measurements using a tunable mid-infrared pulsed laser system. We use an optical parametric amplifier (MIROPA, Hotlight Systems) producing idler wavelengths tunable in the range 2550--4250~nm, pumped by a 1030~nm pulsed laser (Femtolux, Ekspla). The pump delivers 242~fs pulses at a repetition rate of 1.49~MHz. The resonators are illuminated with a focused, linearly polarized beam using a 0.71 NA aspheric lens (Thorlabs C093TME‑E; f = 3~mm; high transparency across the excitation range), giving rise to a spot size of $\approx$4.8~$\mu$m at 3250~nm. Forward and backward excitations are implemented by flipping the sample inside the setup. We maintain a constant average power across the entire excitation range of wavelengths at the position of the sample. This allows for quantitative comparison of HHG signals across the excitation wavelengths range without normalization with respect to the excitation power. 

To focus the laser beam on a single resonator, we image the excitation spot in reflection using the same lens. The reflected light is directed by a mid-IR beamsplitter onto a NIT Tachyon 16k camera equipped with a CaF$_2$ lens (f = 100~mm). We use mid-IR image in reflection for coarse-tuning the positions of the sub-wavelength resonators, and we fine-tune the alignment between the laser focus and the resonator position by monitoring the signal of optical harmonics versus resonator coordinates in the forward direction. The generated harmonics are collected in transmission using a Mitutoyo Plan Apo NIR 100$\times$/0.7~NA objective and are imaged with cameras as well as measured with spectrometers. For the imaging, we use Xenics Bobcat 320 (InGaAs-based detector) for the third harmonic and Starlight Express Trius Pro-H694 Mono for the fifth and seventh harmonics paired with suitable filters (longpass + shortpass). Spectra are detected with a Peltier-cooled Ocean Optics NIRQuest (InGaAs-based detector, 900$-$1700 nm) and a Teledyne Princeton Instruments spectrometer (HRS-750 spectrograph with a Peltier-cooled Pixis 100/400 detector; detection range from 200$-$1180 nm for 300 nm blaze and 150 grooves/mm grating).

Fig.~\ref{fig:SEM}c shows a representative spectrum of the observed third, fifth, and seventh harmonics from a subwavelength resonator. Only odd-order harmonics are observed, which can be expected from bulk nonlinearities of centrosymmetric Si and Si$_3$N$_4$. We note that while centrosymmetric solids structured at the nanoscale may produce strong even-order harmonics \cite{tonkaev2024even} in cases of geometrical symmetry breakage, our design here does not target enhancement of even harmonics, and we correspondingly do not detect signals at even harmonics frequencies.

To systematically investigate the nonlinear response, we perform our measurements over a range of sample diameters from 1.77 to 2.20~$\mu$m for multiple harmonics and pump wavelengths. 
At incident peak power densities of $290$ GW/cm$^2$, the resonantly enhanced volume-averaged intensity was evaluated using the electric-field enhancement factor,
$F_V = \frac{1}{V}\int_V \frac{|\mathbf{E}(\mathbf r)|^2}{|\mathbf{E_0}|^2}\, dV$, where $\mathbf{E}(\mathbf{r})$ is the local electric field, $\mathbf{E_0}$ is the incident field amplitude, and $I_0$ is the incident peak intensity. This yields the average power density inside the resonator  as $I_{\mathrm{dep}} = F_V I_0 \approx 3.47\, I_0 \approx 1~\mathrm{TW/cm^2}$. Furthermore, the localized peak intensity within the resonator was estimated from the maximum electric-field enhancement, defined as $I_{\mathrm{max}}=\frac{\max |\mathbf{E}(\mathbf{r})|^2}{|\mathbf{E_0}|^2} I_0$. Using this definition, the maximum hotspot intensity was found to be approximately $7.67~\mathrm{TW/cm^2}$.
For comparison, bulk silicon exhibits a lower single-pulse damage threshold fluence of approximately $0.5~\mathrm{J/cm^2}$ for $300~\mathrm{fs}$ pulses at 1030 nm wavelengths, corresponding to a peak intensity of roughly $1.67~\mathrm{TW/cm^2}$\cite{prada2024influence}. However, silicon absorption decreases substantially in the mid-infrared spectral region, where our experiments are performed, compared to the 1030 nm wavelength, which falls into Si absorption band with photon energy exceeding the band-gap energy. Therefore, a considerably higher damage threshold can be expected. Also, the laser-induced damage threshold of $\beta$-Si$_3$N$_4$ at $780~\mathrm{nm}$ has been reported to be $5.2\times10^{4}~\mathrm{J/m^2}$ at 220 fs pulse width, corresponding to a peak intensity of approximately $23.6~\mathrm{TW/cm^2}$\cite{zhang2009interaction}. Taken together, these considerations suggest that the fabricated Si--Si$_3$N$_4$ resonators are capable of sustaining the high localized intensities generated under resonant excitation without immediate optical damage. Nevertheless, under prolonged exposure at high excitation powers, gradual material degradation and eventual ablation of the resonator can still occur. This behavior likely arises because, as the local temperature exceeds the materials' deposition temperature ($300^{\circ}$C), the films undergo hydrogen desorption or partial crystallization, leading to irreversible structural change \cite{Jorstad2018,Dani2024}. Consequently, all measurements were performed at an average excitation power of 10~mW, resulting in an incident peak power density of $\approx$290 GW/cm$^2$. 

Figs.~\ref{fig:SEM}d-h show selected examples of high contrasts of the harmonics generated for forward vs backward excitations. Specifically, Figs.~\ref{fig:SEM}d,f,h show examples of third, fifth and seventh harmonics being generated more efficiently under the forward excitation, while Figs.~\ref{fig:SEM}e,g show complementary examples of the opposite scenario of harmonics being generated more efficiently for backward excitation for different diameters and excitation wavelengths.

Fig.~\ref{fig:SEM}j summarizes high forward-to-backward and backward-to-forward contrasts that were observed across the range of resonator sizes and excitation wavelengths.

\begin{figure*}[htb]
    \centering
    \includegraphics[width=1\textwidth]{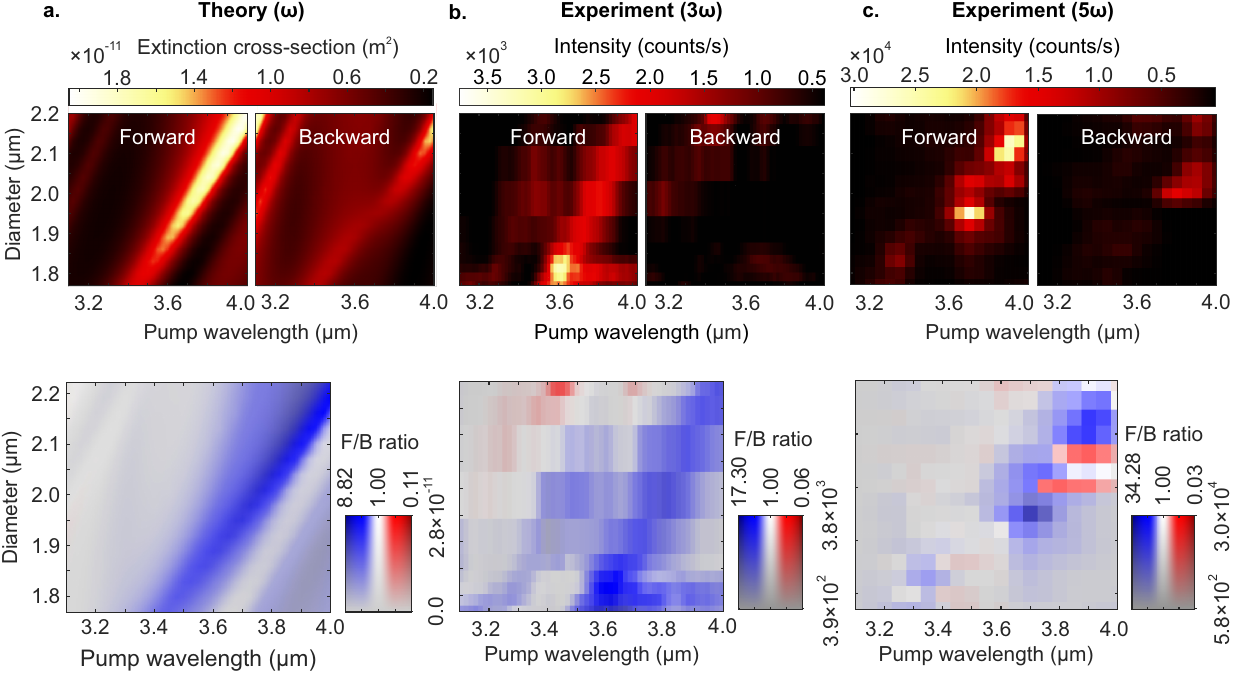}
    \caption{\textbf{Response of asymmetric resonators as a function of excitation wavelength and the resonator diameter.}
    \textbf{(a)} COMSOL calculations of extinction for forward (top left) and backward (top right) directions, as well as forward/backward extinction contrast. In the contrast map, the saturation (from gray to colorful) indicates combined extinction (forward + backward).
    \textbf{(b,c)} Experimental measurements of third-harmonic \textbf{(b)} and fifth-harmonic \textbf{(c)} intensities for forward and backward excitation, together with the corresponding forward/backward intensity contrasts. In the contrast map, the saturation (from gray to colorful) indicates the combined intensity of the generated harmonic (forward + backward) with gray color corresponding to an overall low/undetectable harmonics signal for both directions of excitation.
   }
    \label{fig:hhg_maps}
\end{figure*}

In Figs.~\ref{fig:hhg_maps}b,c, we present systematic measurements of third and fifth harmonics for forward and backward excitation versus two experimental parameters: excitation wavelength and resonator diameter. We made the colorplot for third and fifth harmonics using the Guassian weighted average for a window size of 3$\times$3. We do not perform such a study for the seventh harmonic, as it was observable only for optimal conditions in our experiments and was not detectable outside of resonant geometries and excitation wavelengths. Bottom figures in (b,c) present the corresponding forward-to-backward contrast map, defined as $\mathcal{C} = F/B$, where $F$ and $B$ denote forward and backward harmonic intensities, respectively. In the color-coded plots, blue regions correspond to $\mathcal{C} > 1$, where the forward harmonic generation dominates, whereas the red regions ($\mathcal{C} < 1$) indicate stronger backward generation. Our colormap for the contrast plots has a second dimension, which is color saturation that indicates the overall brightness of the signal (F + B), with gray color corresponding to regions with no signal detected -- neither for forward nor for backward directions. The saturation dimension thus renders in gray color noise divided by noise (e.g., "forward" noise by "backward" noise), thus disregarding such data points.

We observe the maximum $F/B$ contrast ratio for the third harmonic at diameter of 1780 nm and at a pump wavelength of 3605 nm. Similarly, the maximum $B/F$ third harmonic contrast ratio is obtained for a diameter 2140 nm and a pump wavelength of 3184 nm. The maximum $F/B$ contrast obtained is 30, and the maximum $B/F$ contrast obtained is 12.


The maximum fifth harmonic $F/B$ contrast is obtained for a resonator of diameter 1950 nm and for a pump wavelength 3700 nm (Contrast value 55). Similarly, the maximum fifth harmonic $B/F$ value is observed for a resonator of diameter 2060 nm and 3900 nm pump wavelength (Contrast of 13.3). Furthermore, we find pronounced contrast features under forward and backward excitations near the resonant region for pump wavelengths between 3500~nm and 4000~nm in both third- and fifth-harmonic generations, agreeing with the asymmetric near-field distribution predicted by theory. We note that the region of maximum forward intensity for the third harmonic, centred around 3600 nm pump wavelength, does not coincide with that of the fifth harmonic. The same is reflected in the corresponding contrast maps. This discrepancy we attribute possibly due to the different structure of Mie resonances at the wavelengths of the 3$^{rd}$ and the 5$^{th}$ harmonics, which is also influenced by different spectral overlap of the bandgap absorption of amorphous silicon and the fifth-harmonic vs third-harmonic wavelengths\cite{Poruba2004}.


\begin{figure*}[htb]
    \centering
    \includegraphics[width=1\textwidth]{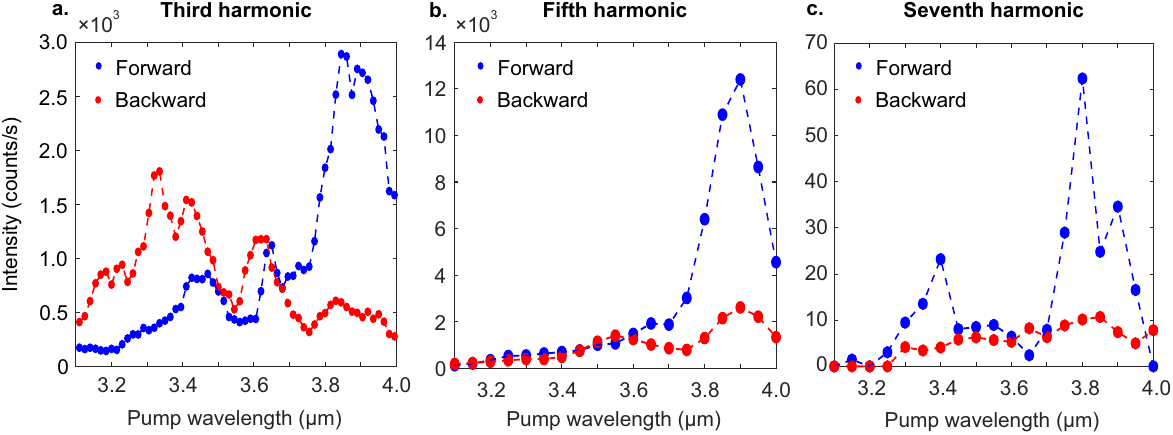}
      \caption{
      \textbf{Asymmetric high-harmonic generation.} Forward (blue) and backward (red) \textbf{(a)} third, \textbf{(b)} fifth, and \textbf{(c)} seventh harmonics generation from the resonator of diameter 2.1 $\mu$m. The third harmonic data were collected using the InGaAs-based detector, while the fifth and seventh harmonic data were collected using the Si-based detector.}
    \label{fig:hhg_fifth}
\end{figure*}

Fig.~\ref{fig:hhg_fifth} presents a comparison of the resonant features emerging in the third, fifth, and seventh harmonic generations for the same resonator with a diameter of 2.1 $\mu$m under both forward and backward illumination. The quantitative analysis of harmonic intensity as a function of pump wavelength shows resonant enhancement across all orders in the vicinity of the numerically predicted regions (see Fig.~\ref{fig:hhg_maps}a). However, backward generation of the fifth and seventh harmonics is suppressed in regions where the third harmonic is present (also see Fig. \ref{fig:hhg_maps}). Notably, the seventh harmonic is not observed across most of the diameters, and a resonator diameter of 2.1 $\mu$m is chosen here as a strong resonant point for the 7$^{th}$ harmonic. 



\section{Conclusion}
In conclusion, we demonstrate asymmetric generation of resonantly enhanced third-, fifth- and seventh-harmonic signals from a bilayer subwavelength resonator. 
The axially asymmetric bilayer configuration of the subwavelength resonator enables a strong dependence of HHG on the directionality of excitation, yielding distinct intensities for the ``forward'' and ``backward'' directions. Our theoretical studies demonstrate the underlying mechanism for the asymmetric harmonic generation arises from the multipolar bianisotropy associated with the $E_1$ irreducible representation of  $C_{\infty v}$ point symmetry group. High-harmonic generation in nanostructured solids opens a route towards compact sources of attosecond pulses and extreme-ultraviolet light \cite{Sarantseva2020_AttosecondMetrology,Johnson2018_SoftXrayHHG,Fu2020_WaterWindowSoftXray}. Our results add a new tool to the toolbox of nanophotonic approaches to both enhance and control the generation of high harmonics in solids using nanotechnology and physics of Mie resonances.


\section*{Acknowledgements}
 We thank Y. Kivshar for his valuable suggestions. We also thank I. Toftul and D. Smirnova for their numerous insightful discussions. The authors acknowledge support from the Australia-Germany Joint Research Cooperation Scheme (57692192) and the Deutsche Forschungsgemeinschaft (DFG, German Research Foundation)-TRR142/3-2022-No. 231447078-projects B09/A08.

\section*{Supplementary Information}
See Supplementary Information for supporting content.

\newpage

\bibliography{bibliography}
\pagenumbering{gobble}

\clearpage
\include{Supplementary_main.tex}

\end{document}

%% file: Supplementary_main.tex
\clearpage

\onecolumngrid

\begin{center}

{\Large\bfseries Supplementary Information}\\[1em]

{\large\bfseries Asymmetric high-harmonic generation from subwavelength resonators}\\[2em]

{\normalsize
Albert Mathew$^{1,*}$,
Piyush Jangid$^{1,*}$,
Rebecca Aschwanden$^{2}$,
Yves Köppeler$^{2}$,
Thomas Zentgraf$^{2}$,
and Sergey Kruk$^{3}$
}

\vspace{1em}

\small
$^{1}$Nonlinear Physics Centre, Research School of Physics,\\
Australian National University, Canberra ACT 2601, Australia

\vspace{0.5em}

$^{2}$Institute for Photonic Quantum Systems (PhoQS) and Department of Physics,\\
Paderborn University, Warburger Straße 100, 33098 Paderborn, Germany

\vspace{0.5em}

$^{3}$Photonics Laboratory, Physics Unit,\\
Tampere University, Tampere, Finland

\vspace{1em}

$^{*}$These authors contributed equally to this work.

\vspace{2em}

\end{center}

\maketitle  
\thispagestyle{empty}
\newpage
\section{Scattering and absorption cross section of the resonator}
\begin{figure*}[htb]
    \centering
    \includegraphics[width=1\textwidth]{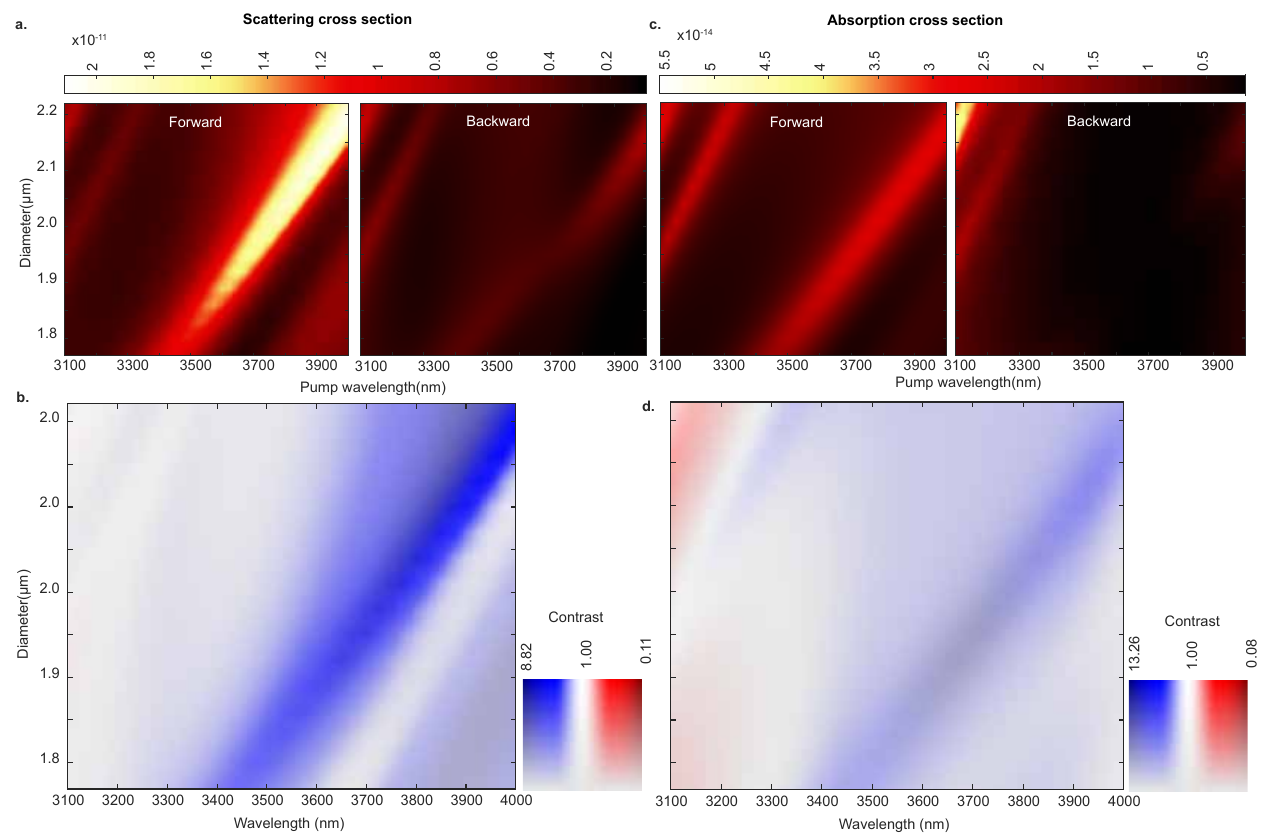}
    \caption{\textbf{Scattering and absorption cross-section} The scattering and absorption cross section of Si-SiN subwavelength resonators simulated using COMSOL as a function of excitation wavelength and the diameter of the cylidrical resonators\textbf{(a)} The colourplot of forward and backward scattering cross section of Si-SiN resonators. \textbf{(b)} The corresponding forward to backward contrast of scattering cross-section obtained from Figure S1.a. \textbf{(c)} The colourplot of forward backward absorption cross-section. \textbf{(d)} The corresponding forward to backward contrast of absorption cross section obtained from Figure S1.c.  }
    \label{fig:concept}
\end{figure*}

\section{Finding Mode volume}
Mode volume is a fundamental parameter that characterizes the spatial confinement of electromagnetic energy and quantifies the strength of light matter interaction within an optical cavity. In this work, we determined the mode volume using finite element simulations performed in COMSOL Multiphysics. The effective mode volume was calculated using

\begin{equation}
V_{\mathrm{mode}} = 
\frac{\int \varepsilon(\mathbf{r}) |\mathbf{E}(\mathbf{r})|^2 \, dV}
{\max\left[\varepsilon(\mathbf{r}) |\mathbf{E}(\mathbf{r})|^2\right]},
\end{equation}

where $\varepsilon(\mathbf{r})$ is dielectric permittivity of the material and $\mathbf{E}(\mathbf{r})$ represents the electric field distribution of the resonant mode across the resonator\cite{alaeian2020cavity}\cite{srinivasan2006cavity}. The quantity $\varepsilon(\mathbf{r}) |\mathbf{E}(\mathbf{r})|^2$ corresponds to the time averaged electromagnetic energy density. In the equation, the numerator represents the total stored electromagnetic energy obtained by integrating the energy density over the resonator and the surrounding near-field region where the electric field is localized.

The calculated mode volumes are reported in normalized units of $\lambda^3$, enabling direct comparison of optical confinement relative to the diffraction limit. For a resonator with a diameter of 2.1~$\mu$m excited at a pump wavelength of 3875~nm, we obtained mode volume as $0.03\lambda^3$, indicating strong subwavelength confinement of the electromagnetic field and enhanced light matter interaction.